\documentclass[preprint,10pt,a4paper]{aastex}
\usepackage{amsmath}
\usepackage{graphicx}
\usepackage{multicol}
\newcommand{\be}{\begin{equation}} 
\newcommand{\ee}{\end{equation}}
\newcommand{\nn}{\mbox{} \nonumber \\ \mbox{} }
\newcommand{\ba}{\begin{eqnarray}}
\newcommand{\ea}{\end{eqnarray}}
\newcommand{\om}{\omega}

\newcommand\eg{\textit{e.g.,\ }}
\newcommand\cf{\textit{cf.\ }}
\newcommand\ie{\textit{i.e.,\ }}

\newcommand{\Bf}{{magnetic field}}

\newcommand{\Ef}{{electric  field}}
\newcommand{\Efs}{{electric fields}}

\newcommand{\ms}{magnetosphere}
\newcommand{\mss}{magnetospheres}
\newcommand{\Fermi}{{\it Fermi}}
\newcommand{\LC}{light cylinder}

\begin{document}

\title{Inverse Compton model of pulsar high energy emission}
\author{Maxim Lyutikov\\
Department of Physics, Purdue University, 
 525 Northwestern Avenue,
West Lafayette, IN
47907-2036, USA
\\
and
\\
INAF - Osservatorio Astrofisico di Arcetri, 
Largo Enrico Fermi 5, I - 50125 Firenze,  Italia }

\begin{abstract} 
We reproduce  the broadband  spectrum of  Crab pulsar, from UV to very high energy gamma-rays - nearly ten decades in energy, within the framework of  the cyclotron-self-Compton model. Emission is produced  by  two counter-streaming beams within the outer gaps, at distances above  $\sim$ 20 NS radii. The outward moving beam  produces UV-$X$-ray photons via Doppler-booster cyclotron emission, and GeV photons  by Compton scattering the cyclotron photons produced by the inward going beam. The scattering occurs in the deep Klein-Nishina regime, whereby the  IC component provides a direct measurement of particle distribution within the magnetosphere.   The required plasma multiplicity is high, $\sim 10^6-10^7$, but is consistent with the average particle flux injected into the pulsar wind nebula. 

 The importance of  Compton scattering in the Klein-Nishina regime also implies the importance of  pair production in the outer gaps. We suggest that outer gaps are important sources  of pairs  in pulsar magnetospheres.

Cyclotron motion of particles  in the pulsar magnetosphere may be excited due to  coherent  emission of radio waves by  streaming particles at the anomalous cyclotron resonance. Thus, a whole range of  Crab non-thermal emission, from coherent radio waves to very high energy $\gamma$-rays - nearly eighteen  decades in energy - may be a manifestation of inter-dependent  radiation processes. 
 
 The present model, together with the observational evidence in favor of the IC scattering (Lyutikov et al. 2012; Lyutikov 2012), demonstrates that the inverse Compton scattering can be the dominant high energy emission mechanism in majority of pulsars.

\end{abstract}

\section{Introduction}

The pulsar high energy emission is a complicated  unsolved problem in high energy astrophysics. It has been  been under intensive study for nearly four decades \citep{cr77,1986ApJ...300..500C,1995ApJ...438..314R,1996ApJ...458..278D,2008ApJ...680.1378H}. The   \Fermi\ Gamma-Ray Space Telescope detected  a large number of pulsars \citep{2010ApJS..187..460A}; this  revolutionized our picture of the non-thermal emission from pulsars in the gamma-ray band from 100\,MeV up to about 10\,GeV. At even higher energies, in the very-high energy (VHE) band, the detection of the Crab pulsar at 25\,GeV by the Magic Collaboration  \citep{2008Sci...322.1221A} and recently at 120\,GeV by the VERITAS Collaboration  \citep{VERITASPSRDetection} in the very-high energy  band  allow to stringently constrain the very-high-energy emission mechanisms

 {\it Geometrical} models, based on the idea of the outer gap \citep{1986ApJ...300..500C}, are very successful in explaining the basic features of the observed $\gamma$-ray light curves \cite[\eg][]{1995ApJ...438..314R,2008ApJ...680.1378H,2010ApJ...715.1282B}. While there seems broad consensus that the particle accelerator is located in the outer magnetosphere, the radiation physics remain controversial \citep[\eg][]{1996A&AS..120C..49A}.

Nearly universally, the origin of the emission above $\sim $ 100 MeV was until recently attributed to the curvature emission  \citep[\eg][]{cr77,1986ApJ...300..500C,1995ApJ...438..314R,1999MNRAS.308...54H,2003ApJ...591..334H,2007ApJ...662.1173H,2008ApJ...680.1378H}. For example, \cite{1986ApJ...300..522C}  concluded that "Crab primary outer gap $e^+/e^-$ lose most of their energy to curvature $\gamma$-rays". Curvature radiation has remained as the preferred gamma-ray emission mechanism \citep{1996ApJ...470..469R} (see also \cite{1986ApJ...300..500C,2000ApJ...537..964C,2008MNRAS.386..748T,2008ApJ...676..562T}). Possible importance of the IC scattering was discussed in application to the Vela pulsar \citep[\eg][]{1986ApJ...300..522C,1996ApJ...470..469R} \citep[see also][]{1986ApJ...300..500C,2000ApJ...537..964C,2008ApJ...680.1378H}. The IC scattering was assumed to be done by the particles in the \ms\  interacting with surface $X$-ray photons. The IC scattering   was not deemed to be {\it the dominant mechanism of high energy emission}. 

\begin{figure}[h!]
\includegraphics[width=.99\linewidth]{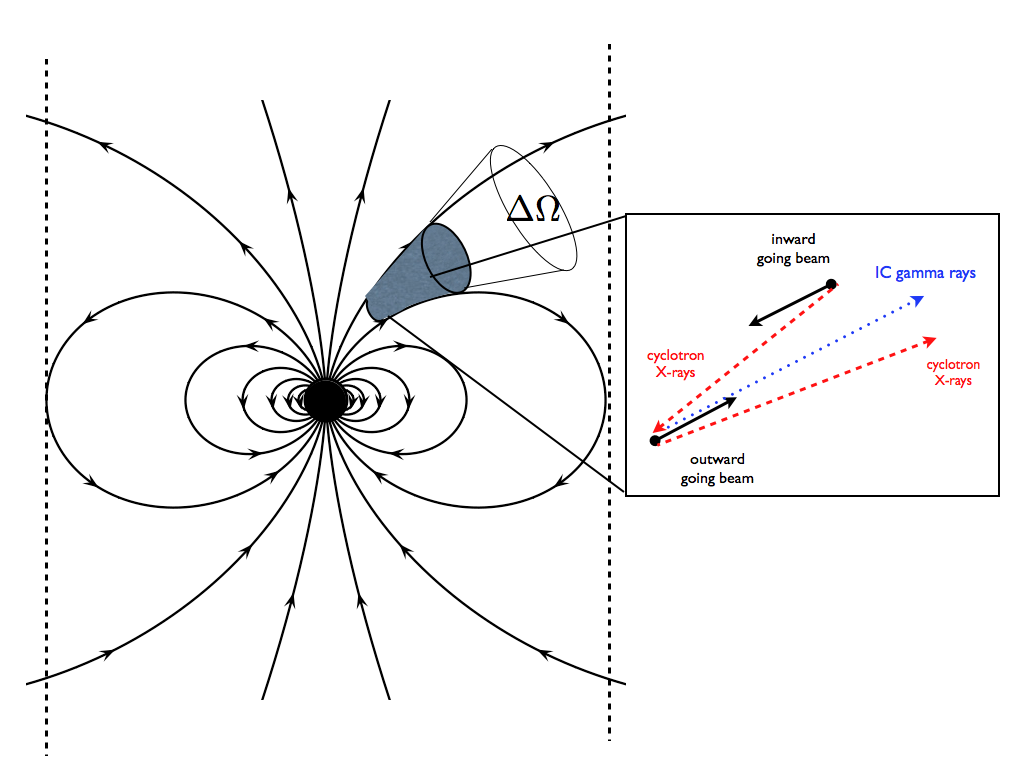}
\caption{Overall geometry. It is assumed that the emission region occupies a region of open field lines with solid opening angle  $\sim  \Delta \Omega$. The insert shows the  radiative processes: two counter streaming beams produce Doppler-boosted cyclotron emission and IC emission on the soft photons of the counter-streaming beam.   } 
\label{picture}
\end{figure}

 In contrast, 
 we argued   \citep{2012ApJ...754...33L,2012arXiv1203.1860L} that  the  IC scattering may be the dominant source of high energy photons in a majority of pulsars \citep[see also][]{2012A&A...540A..69A}. In this paper we further develop the IC model to include a modeling of the broadband SED, from UV to very high energy $\gamma$-rays, covering nearly ten decades in energy. In its essence,  the lower energy UV-$X$-ray peak is due to the cyclotron emission by the secondary particles, Doppler boosted by the parallel motion of the plasma to the $X$-ray range, while the GeV component is due to the scattering of the cyclotron photons by the counter-streaming beam, Fig.  \ref{picture}.

\section{Observed spectrum of Crab}

Crab pulsar produces  non-thermal radiation from radio to very high energy $\gamma$-rays. Radio emission is coherent and has a different origin, though it can be related  to (or actually trigger)  the high energy emission, see \S \ref{anomalos}. The non-coherent non-thermal emission then spans energies from optical, $\sim 1 $ eV,  to very high energy $\gamma$-rays, $\sim 10^{11}$ eV.  We interpret  the Crab SED as having {\it two spectral bumps}  \citep[\eg][and Fig. \ref{CrabPulsarFit}]{2001A&A...378..918K}, a broad UV-$X$-ray-soft $\gamma$-ray bump, $\sim 1 $ eV- $10$ MeV, and a high energy $\gamma $-ray bump, $\sim 100$ MeV- $100$ GeV.  In this paper we address the  nature of the non-coherent non-thermal emission and reproduce the high energy spectrum over nearly 10 decades in energy. 

The low energy part of the  Crab pulsar spectrum may be roughly represented as a $\nu F_\nu \propto \epsilon^{1/3}$ for energies below $\sim 10 keV$, a flat part between $10$ keV- $1$ MeV and a falling $\nu F_\nu \propto \epsilon^{-1/3}$ above MeV \citep{2001A&A...378..918K}. In the high energy band, the spectrum is nearly flat  $\nu F_\nu \propto \epsilon^0$ below few GeVs \citep{2010ApJ...708.1254A}  and shows a long powerlaw tails $\nu F_\nu \propto \epsilon^{-3.8}$ up to $\sim 100 $ GeV \citep{VERITASPSRDetection}.

\section{Outline of the model}
\label{outline}

We assume that emission occurs in the outer gaps with  a conical shape region extending between some minimal radius $r_ {min}$
 and the light cylinder.
The conical shape is naturally an approximation, based on the notion that the high energy emission is produced along caustics, where effects of light travel and aberration compensate. We assume that the emission region subtends a solid angle $\Delta \Omega$.  We assume that the gaps have {\it two} counter-streaming population of secondary particles,  each with multiplicity $\lambda$ and  constant bulk Doppler factor $\delta$.
The presence of counter-streaming populations of leptons is an important ingredient of the outer gap models, \eg \cite{1986ApJ...300..500C} write "Outer gap models will generally result in roughly symmetric flowing streams of relativistic [pairs] in both directions along {\bf B} field lines within the gap". \citep[][also argued that  half of the charged particles produced gaps  move toward the star.]{1993ApJ...415..286H,1998ApJ...493L..35C}
 
In this Section  we also assume that particles have constant transverse velocity $\beta_0$. Later, in \S \ref{opo}, in calculating the SED this assumption is relaxed.  Note that the assumption of constant bulk Doppler factor and constant transverse velocity imply violation of the first adiabatic invariant, which would lead to parallel deceleration and increasing transverse velocity for  incoming beam and parallel acceleration and decreasing transverse velocity for outgoing beam.
The implicit assumption here is that there is an external mechanism that leads to the violation of the first adiabatic invariant, see \S \ref{anomalos}.
Thus, we assume that both beams have equal Lorentz factors $\Gamma$ and equal densities - theses assumptions can easily be relaxed in the follow-up studies.
We also assume that we see emission only from the outward propagating beam: the direct cyclotron emission and the IC upscattering of the {\it inward propagating cyclotron photons}, emitted by the  inward propagating population.

Both inward and outward going  beams produce cyclotron emission - the cyclotron photons emitted by the outgoing beam then produce the UV-$X$-ray bump. 
The  cyclotron photons emitted by the ingoing beam are IC  scattered by the outgoing beam producing the GeV emission. Hence, the model can be called cyclotron-self-Compton, CSC below.  In this Section  we approximate the broad observed distributions of UV-$X$-ray and GeV bumps as having typical values $\epsilon _s$  and $\epsilon _{IC}$. 
Thus, we have four observed quantities (typical energies and fluxes of the cyclotron and IC emission). The unknown quantities are the location of the emission zone (local cyclotron frequency), the relative Lorentz factor of the two plasma components, their density, and the characteristic transverse velocity (which determines the intensity of the cyclotron emission). Thus there are four observables and four unknowns. (Additional parameters are the surface \Bf, determined from pulsar spindown, and the beaming solid angle determined from the pulse profile.)
Thus, neglecting for now  the detailed shape of the spectrums, 
the model has four parameters to be fitted,  $\lambda, r_ {min}, \beta_0, \delta$. This is done using  the typical energies and fluxes of the cyclotron and IC bumps .
These simplifying assumptions are further relaxed in \S \ref{opo}.

\subsection{Kinematics of Compton scattering from counter-streaming beams}

Let us derive the kinematic properties of the cyclotron emission  and IC scattering for a system of two counter-propagating beams. We assume that the IC scattering is done by the outward propagating particles on the cyclotron  photons produced by the inward propagating population. As a first step, we assume that there is a typical frequency of the cyclotron and IC emission, neglecting the fact that both spectral distributions are broad range. This simplification allows us to determine the overall properties of the particles' distribution function required to explain the observations

Let  $\tilde{\epsilon} $ denote  the photon energy measured in the observer frame (the center of momentum frame)  in terms of the electron rest mass energy $\tilde{\epsilon}=\epsilon/( m_e c^2)$. The cyclotron energy  is then measured in terms of the quantum field, $\hbar \om_B  \rightarrow b$, where $b(r)=B/B_Q$,  $B_Q= m_e^2 c^2 /(e \hbar)$.

The forward boosted cyclotron energy is 
\be
\tilde{\epsilon} _s = \delta b  \rightarrow 2 \Gamma b
\label{epsilon _s}
\ee

For the backward propagating photon, its energy in the frame of the 
forward propagating electron is 
\be 
\tilde{\epsilon} _{ERF} = \delta^2 b   \rightarrow 4  \Gamma^2 b
\ee
where $\delta = \sqrt{(1+\beta)/(1-\beta)}$ is the Doppler factor.
This will give an IC photon    with energy in the lab frame
\be
\tilde{\epsilon}_{IC} = \delta { \tilde{\epsilon} _{ERF}  \over 1+ 2 \tilde{\epsilon} _{ERF} }   \rightarrow {8 \Gamma^3 b \over 1+8 \Gamma^2 b},
\label{epsilonIC}
\ee
where the last relations assume $\Gamma \gg 1$.

The KN transition corresponds to $8 \Gamma_{KN}^2 b \sim 1$, 
\be
\Gamma_{KN}  =  {1\over \sqrt{ 8 b} }  = {1 \over 2 \sqrt{2}} b_{NS}^{-1/2} \eta_R ^{3/2} 
\label{KNN}
\ee
where, $b_{NS}=B_{NS}/B_Q $; for dipolar \Bf\ $b=b_{NS}/\eta_R ^3$, 
 and $\eta_R= r/R_{NS}$.

Eqns  (\ref{epsilon _s}-\ref{epsilonIC}) can be resolved for the local \Bf\  at the   location of the initial emission of the cyclotron photon and the Doppler factor:
\ba &&
b=\tilde{\epsilon} _s^2 \left( \sqrt{1+{1\over \tilde{\epsilon} _s \tilde{\epsilon} _{IC}}} -1\right) \approx
\left\{
\begin{array}{cc}
{\tilde{\epsilon} _s^{3/2} /  \tilde{\epsilon} _{IC}^{1/2}}, & \tilde{\epsilon} _s \tilde{\epsilon} _{IC} \rightarrow 0 \\
{\tilde{\epsilon} _s /( 2 \tilde{\epsilon} _{IC})},  & \tilde{\epsilon} _s \tilde{\epsilon} _{IC} \rightarrow \infty
\end{array}
\right.
\nn &&
\delta  = \tilde{\epsilon} _{IC} \left( \sqrt{1+{1\over \tilde{\epsilon} _s \tilde{\epsilon} _{IC}}} +1 \right)
 \approx
\left\{
\begin{array}{cc}
\sqrt{  \tilde{\epsilon} _{IC} / \tilde{\epsilon} _s},  & \tilde{\epsilon} _s \tilde{\epsilon} _{IC} \rightarrow 0 \\ 
2  \tilde{\epsilon} _{IC},  & \tilde{\epsilon} _s \tilde{\epsilon} _{IC} \rightarrow \infty
\end{array}
\right.
\label{22}
\ea
where the two limits correspond to Thomson and KN regimes.

The rest-frame photon energy is then
\be
\tilde{\epsilon} _{ERF}=  \tilde{\epsilon} _{IC} \tilde{\epsilon} _s \left( \sqrt{1+{1\over \tilde{\epsilon} _s \tilde{\epsilon} _{IC}}} +1\right) \approx
\left\{
\begin{array}{cc}
\sqrt{ \tilde{\epsilon} _s  \tilde{\epsilon} _{IC}}, & \tilde{\epsilon} _s \tilde{\epsilon} _{IC} \rightarrow 0 \\
2 \tilde{\epsilon} _s   \tilde{\epsilon} _{IC},  & \tilde{\epsilon} _s \tilde{\epsilon} _{IC} \rightarrow \infty
\end{array}
\right.
\ee

In terms of the surface \Bf\ and stellar radii, Eqns \ref{22} give for the location of the emission zone
\be
\eta_R = {b_{NS} ^{1/3} \over \tilde{\epsilon} _s^{2/3}  \left( \sqrt{1+{1\over \tilde{\epsilon} _s \tilde{\epsilon} _{IC}}} -1\right)^{1/3} }
 \approx
\left\{
\begin{array}{cc}
b_{NS}^{1/3} {\tilde{\epsilon} _{IC}^{1/6} /  \tilde{\epsilon} _{s}^{1/2}}, & \tilde{\epsilon} _s \tilde{\epsilon} _{IC} \rightarrow 0 \\
2 ^{1/3} b_{NS}^{1/3} {\tilde{\epsilon} _{IC}^{1/3} /  \tilde{\epsilon} _{s}^{1/3}},  & \tilde{\epsilon} _s \tilde{\epsilon} _{IC} \rightarrow \infty
\end{array}
\right.
\ee

The Compton cross-section for forward scattering
\ba && 
\sigma= { 1+2 b \delta^2 + 2 b^2 \delta ^4 \over (1+ 2 b \delta^2)^3}{r_E^2}  ={ 1+2  \tilde{\epsilon} _{s} \tilde{\epsilon} _{IC} \over \left(1+  2 \tilde{\epsilon} _{IC} \tilde{\epsilon} _{s}  (1+  \sqrt{1+1/( \tilde{\epsilon} _{IC} \tilde{\epsilon} _{s})}) \right)^2} {r_E^2} =
  { (\delta- 2  \tilde{\epsilon} _{IC}) (\delta^2 - 2 \delta  \tilde{\epsilon} _{IC} + 2  \tilde{\epsilon} _{IC}^2 )\over \delta^3}  {r_E^2}
    \nn &&
\approx
\left\{
\begin{array}{cc}
{r_E^2}   & \tilde{\epsilon} _s \tilde{\epsilon} _{IC} \rightarrow 0 \\
{r_E^2 \over 4 b \delta^2} = {r_E^2 \over 8 \tilde{\epsilon} _{IC} \tilde{\epsilon} _{s}}& \tilde{\epsilon} _s \tilde{\epsilon} _{IC} \rightarrow \infty
\end{array}
\right.
\ea
where $r_E = e^2/(m_e c^2)$ is the classical radius of an electron.
The transition to the KN regime occurs at $ \tilde{\epsilon} _{IC} \approx 1/( 8  \tilde{\epsilon} _{s})$.

Let us next apply these relations to the Crab pulsar. 
For Crab pulsar   the surface \Bf\ $B_{NS} = 4\times 10^{12} $ G $\rightarrow b_{NS} \approx 0.1$. Taking the observed cyclotron peak at $\sim 50$ keV ($\tilde{\epsilon}_s =0.1$), the IC peak at $\sim 1$ GeV ($\tilde{\epsilon}_{IC} =2000$), gives  the  estimates of the minimal  emission height and the bulk Doppler factor.
\ba &&
\eta_R \approx 15
\nn &&
b\approx  {\tilde{\epsilon}_s \over 2 \tilde{\epsilon}_{IC}} = 2.5 \times 10^{-5}
\nn &&
\delta \approx 4000
\label{dfdf}
\ea
(In Crab the \LC\ is located at $\eta_{R, Max}= 160$.)

The KN transition in Crab corresponds to 
\be
\Gamma_{KN}    \approx \eta_R ^{3/2} 
\label{KNN1}
\ee
Since $ 2 \delta^2 b \approx  4 \tilde{\epsilon}_{IC} \tilde{\epsilon}_{2}=800 \gg 1$, the scattering occurs in a deep KN regime.

Both estimates (\ref{dfdf}) are reasonable. Most model of the high energy emission place the emission heights at tens stellar radii (\cite{2001ApJ...554..624H,2004AdSpR..33..552B}, Arons, priv. comm.), while radiative model predict bulk Lorentz factors in thousands \cite[\eg][]{1999MNRAS.308...54H}.

\subsection{Particle density}
We parametrize the total flux through the gap in terms of the total Goldreich-Julian flux through the open field lines,
\be
\dot{N}_{GJ} = {b_{NS} \over 2} { c R_{NS} \over r_E \lambda_C} \eta_\Omega^2,
\ee
($\eta_\Omega= \Omega R_{NS} /c$),
times the multiplicity factor within the emission region $\lambda$, times the relative  opening angle of the gap region at distance $r$, 
$ \Delta \Omega c /( \pi r \Omega) = \Delta \Omega /( \pi \eta_R \eta_\Omega)$,
\be
\dot{N} = \Delta \Omega \eta_\Omega {b_{NS} \over 2 \pi} { \lambda \over \eta_R} { c R_{NS} \over r_E \lambda_C}
\ee
A number density  of particles at a radius $r$ is then
\be
n= {\dot{N} \over c}  dr = {b_{NS} \over 2 \pi} { \lambda \eta_\Omega \Delta \Omega  \over \eta_R} { c R_{NS} \over r_E \lambda_C} dr
\label{nmmn}
\ee
Multiplicity $\lambda$ and location of the emission $\eta_R$ are the model parameters to be fitted.

\subsection{Cyclotron fluxes}
\label{cycc}
 In the electron center-of-gyration frame the total single particle photon emissivity (integrated over emission angles, photons per second per electron) is
\be
\eta_{ph}^{' (sp)} = 
{2 \over 3}  {c r_E \over \lambda_C^2} \beta_0^2 b \delta_D( \epsilon_s' -  b m_e c^2) d \epsilon_s'
\ee
In the observer frame the photon emissivity is $\delta$ times higher
\be
\eta_{ph} ^{(sp)}=\delta\, {2 \over 3}  {c r_E \over \lambda_C^2} \beta_0^2 b \delta_D( \epsilon_s - \delta b m_e c^2) d \epsilon_s
\label{kkk}
\ee
where   $\lambda_C= \hbar / (m_e c)$ is Compton wavelength, $\delta_D$ denotes Dirac delta-function. 

The total cyclotron emissivity is (\ref{kkk}) times the number of emitting particles (\ref{nmmn}):
\be
\eta_{ph} = { 1 \over 3 \pi} { b_Q^2 \beta_0^2\, \delta\,  \Delta\Omega\, \lambda \over \eta_R^4} { c R_{NS} \over \lambda_C^3} \,
dr \delta_D \left(\tilde{\epsilon} - b \delta  \right) d  \tilde{\epsilon} 
\label{etatot}
\ee

Integrating over radius we find the total photon emissivity of a mono-energetic stream of particles:
\be
\eta_{ph}= {1 \over 9 \pi } b_{NS} \lambda \eta_\Omega \Delta \Omega \beta_0^2 {c R_{NS}^2 \over \lambda_C^3} d \tilde{\epsilon}
\label{etaRR}
\ee
Thus a beam propagating in a dipolar \Bf\ produces constant luminosity between the minimal and maximal   limits, 
$ \tilde{\epsilon}_{min}= b_{NS} \delta/\eta_{R, Max}^3$ and
$ \tilde{\epsilon}_{max}(r)= b_{NS} \delta/\eta_R^3$.

The minimal energy $ \tilde{\epsilon}_{min}$ is emitted at the light cylinder, 
\be
 \tilde{\epsilon}_{min}= b_{NS}\, \delta \,\eta_\Omega^3.
 \ee
  The maximal energy at each location $\eta_R$ is 
\be
 \tilde{\epsilon} _{min}=  \delta b_{NS} \eta_{R}^{-3}
 \ee

The total cyclotron luminosity, emitted mostly at the innermost limit of the gap, can then be estimated using Eq. (\ref{etaRR}) as
\be
L_X =  m_e c^2 \int   \tilde{\epsilon}   \eta_{ph}   d \tilde{\epsilon}  \approx {1 \over 18 \pi} b_{NS} \lambda \eta_\Omega \Delta \Omega \beta_0^2 { m_e c^3   R_{NS}^2 \over \lambda_C^3}   \tilde{\epsilon}^2
\label{LS}
\ee

Equation (\ref{LS}), together with the estimate of the peak energy, $  \tilde{\epsilon}= \delta b_{NS} /\eta_{R, min}^3$, gives the two constraints on the parameters of the model ($\lambda, \eta_{R, min}, \beta_0, \delta$). The other two constraints will come from the similar estimates for the IC component. 

\subsection{IC emission}
\label{ICemission}
In \S \ref{cycc} we estimated the cyclotron emission produced by an outgoing flux of particles.
To estimate the IC flux we assume that the same cyclotron flux is produced by the ingoing beam. Then, at each radius IC scattering occurs on the cyclotron photons produced by the ingoing beam between that radius and the maximal radius - the \LC\ radius. The total IC emission is then the integral of emissivities over the emission zone, between minimal and maximal radii. Thus, for a cold flow, a given $\epsilon_{IC}$ comes from a  cyclotron photon 
emitted at given height, but the IC scattering can occur anywhere 
inside.

The maximal IC energy, corresponding to cyclotron emission and IC scattering near the stellar surface, is
\be
\tilde{\epsilon}_{IC} = {\delta^3 b_{NS} \over 1+ 2 b_{NS} \delta^2} \approx \delta/2
\ee
Where the last  relation takes into account that $b_{NS} \approx 0.1$, so that for any $\delta \geq 2$ the scattering near the surface occurs in the KN regime.

The minimal  IC energy, corresponding to cyclotron emission at the maximal $\eta_{R, max}= c \Omega/R_{NS}$ is
\be
\tilde{\epsilon}_{IC} = {\delta^3 b_{NS}/\eta_{R, max}^2  \over 1+ 2 b_{NS} \delta^2/\eta_{R, max}^3}
\ee
If the KN regime dominates even for lowest energy cyclotron photons, all the IC photons come out with
$\tilde{\epsilon}_{IC} \approx \delta/2 \approx \gamma$ independent of the emission height of the target cyclotron photon.

The 
photon emissivity rate  $\eta_{ph,IC}$ per electron is the  number of collisions per unit 
time of soft photons with the electron.
The soft photon density in  the lab frame, at  the location $\eta_R$ is proportional to the emissivity integrated from $\eta_R$ to the $\eta_{R,Max}$, Eq. (\ref{etaRR}),
\be
n_s = {\eta_{ph} \over R_{NS}^2 c \Delta \Omega \eta_R^2}= {b_{NS} \over 9 \pi} \beta_0^2 \eta_\Omega \lambda {1\over \eta_R^2} {1\over \lambda_C^3} 
\label{nnss}
\ee
In the forward moving electron frame, the  density of soft photons emitted 
towards the star   is $\Gamma$ times higher. (Recall that we assume that outgoing and inward going particle and photon fluxes are similar). 
\be
n_s' = \Gamma n_s
\ee
The IC scattering rate in the forward moving plasma frame is then 
\be
\eta_{ph,IC} '= n_s' \sigma c \delta
\ee
In the lab frame it is $\Gamma$ times smaller
\be
\eta_{ph,IC} = \eta_{ph,IC}'/\Gamma= n_s \sigma c
\ee
Thus, the total IC photon luminosity produced at radius $r$ by the outward moving beam of density $\dot{N} dr /c$ is
\be
\eta_{ph,IC} = n_s c \sigma \dot{N} {dr \over c} \delta_D\left(\tilde{\epsilon}_{IC} - { \delta^2 \tilde{\epsilon}_{s} \over 1+ 2 \delta  \tilde{\epsilon}_{s}}\right) d \tilde{\epsilon}_{IC}
\label{dddfg}
\ee

Using photon density (\ref{nnss}), number density (\ref{nmmn}) and
integrating over the soft photon energies, using the relation 
\be
 \delta_D \left( \tilde{\epsilon}_{IC} - { \delta^2  \tilde{\epsilon}  \over 1+ 2 \delta  \tilde{\epsilon} } \right)  d \tilde{\epsilon} =  \delta_D \left( \tilde{\epsilon} - { \tilde{\epsilon}_{IC}  \over \delta (\delta - 2 \tilde{\epsilon}_{IC})}\right) {d \tilde{\epsilon} \over (\delta - 2   \tilde{\epsilon} _{IC})^2},
 \ee
 the IC photon emissivity (photons per second)  becomes
\ba &&
\eta_{IC} = {1\over 36 \pi^2} b_{NS}^2 \beta_0^2   \Delta \Omega \eta_\Omega^2  \lambda^2 {c r_E R_{NS}^2 \over \lambda_C^4} F_{\tilde{\epsilon} _{IC}} d\tilde{\epsilon} _{IC}
\nn &&
F_{\tilde{\epsilon} _{IC}} = { \delta^2 -2 \delta \tilde{\epsilon} _{IC}+ 2 \tilde{\epsilon} _{IC}^2 \over \delta^3 (\delta -2 \tilde{\epsilon} _{IC})  }
\left({1 \over \eta_{R,min}^2 } - {\tilde{\epsilon} _{IC} ^{2/3} \over  b_{NS}^{2/3} \delta^{4/3} (\delta - 2 \tilde{\epsilon} _{IC} )^{2/3} } \right),
\label{klll}
\ea
see Fig. \ref{CrabPulsarFit}.
The maximal IC energy is 
\be
\tilde{\epsilon} _{IC, max}= {b _Q \delta^3 /   \eta_{R,min}  \over 1+ 2 b_{NS} \delta^2 / \eta_{R,min} ^3}
\label{ddd}
\ee

The overall shape of the IC spectrum depends on whether a KN regime is reached or not. Far below the KN limit,  at energies much smaller than the maxim energy (\ref{ddd})
\be
F_{\tilde{\epsilon} _{IC}} \approx {1\over \delta^2 \eta_{R,min}^2 }
\ee
If the KN regime is not reached, the spectral power smoothly goes to zero as $\tilde{\epsilon} _{IC} \rightarrow \tilde{\epsilon} _{IC, max}$. But if the KN regime is  reached somewhere in the \ms, the distribution develops a peak near $\tilde{\epsilon} _{IC}  \sim \delta /2$. The peak is narrower and higher in the deeper KN regime: in this case all the photons come out at the energy of the electron, regardless of their initial energy.

\begin{figure}[h!]
\begin{center}
\includegraphics[width=.99\linewidth]{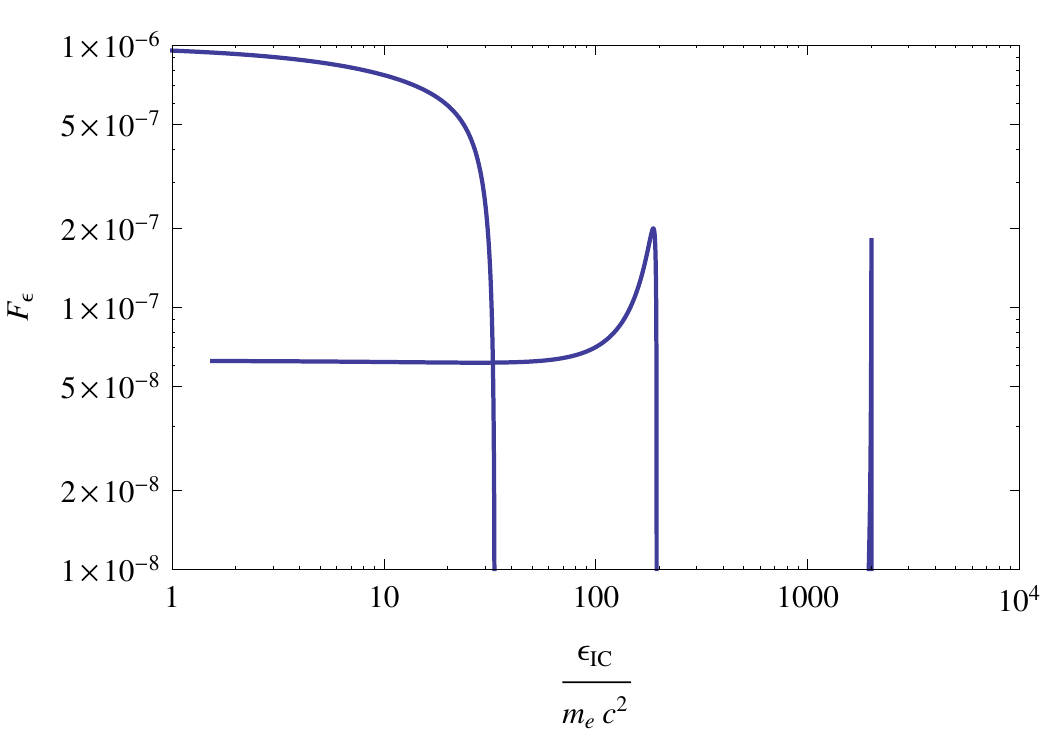}
\caption{Spectrum of IC  photons for mono-energetic beam propagating in the \ms\ and scattering cyclotron photons from a similar inward propagating beam.
Different bulk Doppler-Factors are plotted $\delta= 100, \, 400, \, 4000$. For higher $\delta$, when scattering enters KN regime, the spectrum shows a peak near $\tilde{\epsilon} _{IC}  \sim \delta /2$: in this case  the photons scattered in the KN limit  come out at the energy of the electron, regardless of their initial energy.
}
\end{center}
\label{FepsilonIC2}
\end{figure}
 
 The maximal IC energy is
\be
\tilde{ \epsilon}_{IC, Max}=  b_{NS} {\delta^3 \over \eta_{R, Min}^3 + 2 b_{NS} \delta^2}
\label{klkl1}
\ee

Relations for the soft photon energy and fluxes (\S \ref{cycc}) and for IC luminosity and energy, (\ref{klll}-\ref{klkl1}) can be resolved for the four parameters of the model $\lambda, \eta_{R, min}, \beta_0, \delta$.
Due to the  dependence of the cross-section on energy, the general solutions of such system are complicated. As we demonstrate below, majority of the IC scatterings occur in the deep KN regime. This allows a considerable simplification.

\subsection{Scattering in the deep KN regime}

In the KN limit all the photons come out with the same energy, $\tilde{ \epsilon}_{IC}= \delta/2$. The cyclotron photon is then emitted in a region where $ b = \tilde{ \epsilon}_s /( 2 \tilde{ \epsilon}_{IC}) \rightarrow \eta_{R}  = (  2 b_{NS}\tilde{ \epsilon}_{IC}/\tilde{ \epsilon}_{s} )^{1/3}$.   
 
The IC cross-section in the KN regime is 
\be
\sigma = {r_E^2 \over 4 \tilde{ \epsilon}_s \delta}
\ee
The IC emissivity  is then
\be
\eta_{ph,IC} = {b_{NS}^2 \over 72 \pi^2}  \beta_0^2 \Delta\Omega \eta_\Omega^2 \lambda^2 { c r_E R_{NS}^2 \over \lambda_C^4} {1 \over \delta}  {  d \tilde{\epsilon}_{s}  \over   \tilde{\epsilon}_{s}} { d \eta_R \over \eta_R^3} \delta_D (  \tilde{\epsilon}_{s} -\delta/2 )d  \tilde{\epsilon}_{IC}
\label{klkl22}
\ee
Integration over  radius gives
\be
\int _{ (b_{NS} \delta /\tilde{\epsilon}_s)^{1/3} }^ {\eta_{R,Max}} { d \eta_R \over \eta_R^3} \approx {1\over 2} \left({b_{NS} \delta /\tilde{\epsilon}_s} \right)^{1/3},
\ee
 while integration over  soft photon density gives
 \be
 \int {  d \tilde{\epsilon}_{s}  \over   \tilde{\epsilon}_{s}} = \ln {  \tilde{\epsilon}_{s,Max} \over  \tilde{\epsilon}_{s, Min} } 
  \ee
The IC luminosity is then
\be
L_{IC} = {\delta \over 2} m_e c^2  \eta_{ph,IC}  =
{b_{NS}^2 \over 144 \pi^2}  \beta_0^2 \Delta\Omega \eta_\Omega^2 \lambda^2 { c^3 m_e r_E R_{NS}^2 \over \lambda_C^4} 
 \ln {  \tilde{\epsilon}_{s,Max} \over  \tilde{\epsilon}_{s, Min} } 
\label{klkl23}
\ee

Scaling of the IC luminosity with density squared ($\propto \lambda^2$) reflects the fact that both the target photon density and the IC scattering rate are proportional to density. The logarithm of the ratio of the highest and lowest energies reflects the fact that the soft photon spectral density is constant (see Eq. \ref{etaRR}), while in the KN regime the cross-section decreases with energy $\propto 1/ \tilde{\epsilon}$.

The two expressions for cyclotron (\ref{LS}) and IC luminosity (\ref{klkl23}) can be resolved for the two remaining parameters of the model, the multiplicity $\lambda$ and transverse velocity $\beta_0$. Assuming   $\tilde{\epsilon} \approx 5 \times 10^4 {\rm eV}/(m_e c^2)$ and $ \tilde{\epsilon}_{IC}\approx 10^9 {\rm eV}/(m_e c^2)$ and given luminosities $L_X\approx 10^{35} $ erg s$^{-1}$ and $L_{IC}\approx 10^{35} $ erg s$^{-1}$, we find:
\ba &&
\lambda = {8 \pi \over b_{NS}}  \tilde{\epsilon}_s^2 {\lambda_C \over r_E} {1\over \eta_\Omega} {L_{IC} \over L_X} \left(  \ln  \left(  { \tilde{\epsilon} _{max} \over \tilde{\epsilon}_{min} }\right) \right)^{-1} \approx 10^4
\nn &&
\beta_0 ={ 2 \sqrt{r_E} \lambda_C \sqrt{\ln { \tilde{\epsilon} _{max} \over \tilde{\epsilon}_{min} }} \over
\sqrt{ \pi m_e c^3 \Delta \Omega}   \tilde{\epsilon}^2 L_{IC}^{1/2} }\approx 4 \times  10^{-5} \Delta \Omega_{-2} ^{1/2}
\label{ff}
\ea

We can also verify that the kinetic energy flux, $ \sim (\delta/2) \lambda n_{GJ} m_e c^3 4 \pi r^2 \Delta \Omega$ is much smaller than the spin-down luminosity $L_{SD}$. This requires  
\be
\lambda \delta < {  b_{NS} \eta_{R} \over \delta \Delta \Omega} \, \eta_\Omega^3 \, {R_{NS} \over \lambda_C} = 6 \times 10^{10} \eta_{R}  \Delta \Omega_{-2} ^{-1},
\label{klkpp}
\ee
a condition that is well satisfied for parameters in Eq. (\ref{ff}).

Also, we can verify that  the ratio of the $X$-ray luminosity, Eq. (\ref{LS}) to the  the spin-down luminosity,
\be
{L_X \over L_{SD}}= {b_{NS} \over 32} \beta_0^3 \lambda \delta^2 \Delta \Omega  {r_E \over \lambda _C} \eta_R ^{-6} \eta_\Omega ^{-3} = { L_X \lambda_C ^3 r_E \over b_{NS}^2 m_e c^3 R_{NS}^2 \eta_\Omega^4} = 10^{-4} 
\ee
corresponds approximately to the typical observed values \citep[\eg][]{2006csxs.book..279K}. 

The estimates given in this Section demonstrate that under fairly general assumption about the parameters of the magnetospheric plasma and, most importantly, assuming a presence of two counter-streaming populations of leptons, it is possible to reproduce the overall properties of the pulsar high energy emission. In the following Section we develop a more detailed semi-analytical model, taking into account a broad distribution of parallel momenta for both beams.

\section{Broadband model of Crab SED}
\label{opo}

\subsection{}
In this Section we build a broad-band cyclotron-self-Compton model of the Crab pulsar SED. 
Note, that the conventional expressions for the synchrotron-self-Compton  (SSC) emissivities  that are used, \eg in  studies of blazars \citep[\eg][]{1992MNRAS.258..657C} are not applicable to  pulsar \mss. Typically blazar  SSC models assume isotropic distribution both  of particles and photons  in some given frame (blob rest frame  moving with a single given velocity), tangled \Bf\ and a power law particle distribution function stretching between some minimal and maximal Lorentz factors. All of these assumptions are not applicable to pulsars.

First, in many pulsars (\eg in Crab), the cyclotron decay times even at the light cylinder are shorter than the period. 
(The spontaneous decay times in the particle rest-frame is 
\be
\tau_c' \approx {m_e^3 c^5 \over B^2  e^4}
\label{refff}
\ee
In Crab, for a particle at rest, the  cyclotron decay time is always smaller than a period, $\tau_c$, by a factor $3 \times 10^{-4}$ at the \LC. For a relativistically moving particle $\tau_c $ is Doppler-stretched; in the observer frame $ \tau_c = \gamma \tau_c'$. It becomes of the order of the period at
\be
\eta_{R,c}= { B_{NS}^{1/3} e^{2/3} P^{1/6} \over c^{5/6} \sqrt{m_e} \gamma^{1/6}}  \approx { 300 \over \gamma^{1/6}}
\ee
Thus, at the \LC\ particles with $\gamma \geq 80$ are in the slow decaying regime, while at the inner edge of the emission region, at $\eta_{r, min}=20$,  particles with $\gamma \geq 10^7$ are in the slow decaying regime.) Thus, the {\it  distribution function is expected to the highly anisotropic}, with small pitch angles. Efficient radiative decay may lead to the {\it non-relativistic transverse velocities}, so that emission occurs not in the synchrotron but in the cyclotron regime.

Second, the pulsar \Bf\ is regular dipolar,  changing in its strength over many orders of magnitude within the emission region. Also,  the cyclotron emission depends on the direction to the observer from a given plasma element. Third, and most importantly,  the distribution of parallel momenta with pulsar \ms\ is very broad. Thus, the standard off-the-shelf SSC models are not applicable to pulsar \mss.

\subsection{Distribution function}

The distribution function within the pulsar \ms\  is bound to a complicated, anisotropic function that strongly depends on the location within the \ms. A self-consistent model of the high energy radiation should take into account evolution of the perpendicular and parallel momenta of particles due to the motion in the inhomogeneous \Bf\ (\eg conservation of the first and second adiabatic invariant if applicable), as well as their evolution due to the other EM processes that can excite or de-excite the transgression motion and/or the parallel component of the momentum (\eg excitant of transverse motion due to the anomalous Doppler resonance, see  \cite{Kaz91,lbm99} and \S \ref{anomalos}.
 It is currently beyond our abilities to calculate its details and its evolution within the \ms\  reliably and self-consistently. We can only hope to catch its main properties by appealing to the basic theoretical ideas and using observations to probe it. The theoretical assumption will naturally give only an approximation, hopefully self-consistent within a given model.  The requirement on the model then is that using the minimal number of parameter it should be able to reproduce the overall broadband properties of the pulsar emission. Below we describe such a model. We stress that due to the complicated nature of the problem, the fit parameters we derived for the particle distribution are probably not unique and also, and not precisely determined within this simple model. Yet, the model is able to reproduce the bulk properties of the Crab high energy emission using one major assumption (that of a presence of counter-streaming plasma components) and a number of parameters derived from observations, like the particle spectrum and its evolution with radius. 

As mentioned above, \S \ref{outline}, we assume that the gaps have {\it two} counter-streaming population of secondary particles. This has been a common assumption in many models, \eg \cite{1986ApJ...300..500C}.
We also assume that we see emission only from the outward propagating beam: the direct cyclotron emission and the IC upscattering of the {\it inward propagating cyclotron photons}, emitted by the  inward propagating population. This produces two spectral bumps, in the UV-$X$-ray range, which we would call the low energy bump, and the $>100$ MeV -- GeV feature, which we would call the high energy bump.

\subsubsection{Parallel distribution}

 Different parts of the spectrum  probe different parts of the particle distribution. Since in the KN regime of the IC scattering the energy of the photons is of the same order as the energy of the scattering electron regardless of the target photon energy, the high energy bump provides a direct measurement of the bulk particle population (weighted by the energy dependence of the KN cross-section).  Since above break the observed spectral index is $\leq - 4$ \citep{VERITASPSRDetection}, this required that the particle spectrum above the break is $\leq -3$ (in the deep KN regime). Below the break the the observed spectral index is flat \cite{2010ApJ...708.1254A},  this required that the particle spectrum well below the break is $\sim 1$.  The modeling also then requires that there should be another energy region with   the particle spectrum below the break is $\sim -1$. 
 
 The high energy part of the low energy bump provides another constraint on the distribution function at the highest energies.
Note that if at the highest energies the particle spectrum is a power-law, $f \propto \delta^{-l}$, the spectral energy distribution is  $\propto \epsilon ^2 \eta_s \propto \epsilon ^{3-l}$. Thus, to have a finite emissivity one either needs $l > 4$ or there should be an upper cut-off to the particle distribution.   (For $|l| < 3$ the cyclotron component keeps rising at high energies, contrary to observations). Also, there should be an exponential cut-off to the particle energies to avoid cyclotron component showing up in the VERITAS band. 
\footnote{In this context we note, that in principle {\it all} the pulsar high energy emission, from  $X$-ray to $\gamma$-rays, can be modeled as boosted cyclotron emission with a population of particles with $l \sim 3$.  
The GeV bump is then just a bump in the parallel distribution. 
Variations in the $X$-ray and $\gamma$-ray profiles then can be due to somewhat different spacial distributions of corresponding particles.
We disfavor this possibility since such a model would require highly correlated $X$-ray to $\gamma$-ray signals.}
 
  
     In addition, there should be a minimal  Lorentz factor $\gamma_0$, of the order $\gamma_{min} \sim 1/\sqrt{\Delta \Omega} 
\sim 10$. In this case the observed profile is mostly determined by the geometrical factors and is nearly energy-independent. (In addition, our simplifying assumption of the IC scattering in the deep KN regime is likely to break down for the low energy tails of the distribution function, see Eq. (\ref{KNN}).)

\begin{figure}[h!]
\includegraphics[width=.99\linewidth]{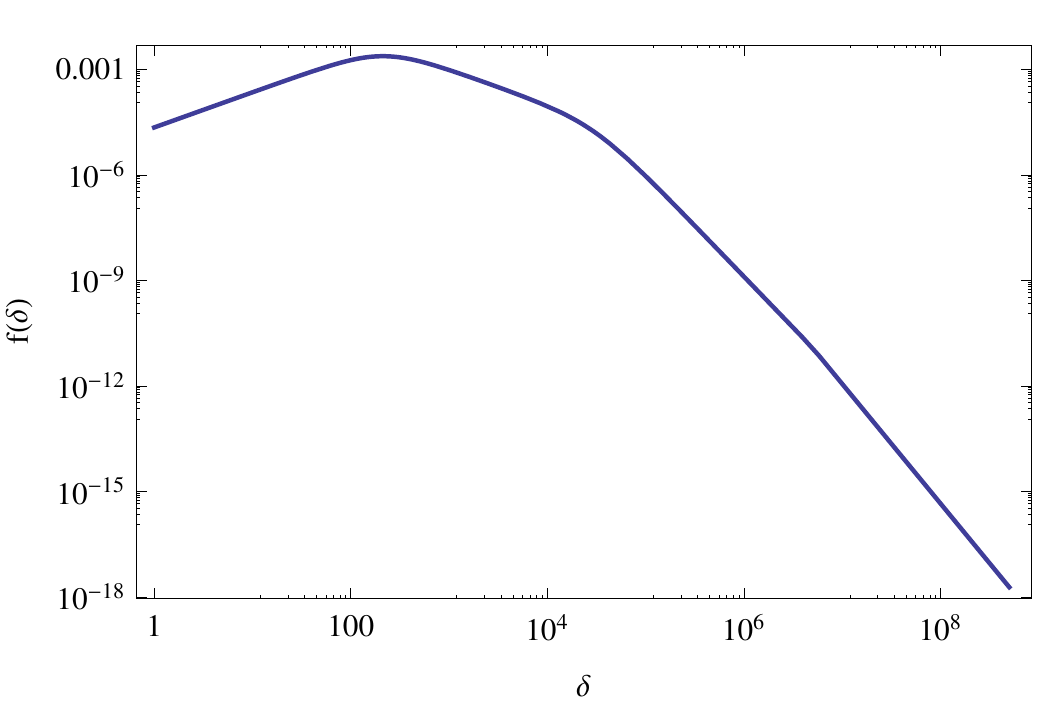}
\caption{The  parallel distribution function $f(\delta)$, Eq. (\ref{ppp}). In addition, there is a counter-streaming beam with the same  distribution. Various parts of the SED constrain various parts of the distribution function; see text for details.} 
\label{fofdelta}
\end{figure}

 Summarizing, the following distribution function of parallel momenta is inferred from observations, see Fig. \ref{fofdelta}:
 \ba &&
 f(\delta) \propto
 \left\{
 \begin{array}{cc}
 \delta ^ m, &  m \approx 1,\mbox{ for } \delta_0 < \delta < \delta_1   \\
 \delta ^ n, &  n \approx -1, \mbox{ for }  \delta_1< \delta < \delta_2  \\
 \delta ^l, & l  \approx -3, \mbox{ for }  \delta_2< \delta < \delta_3  \\
 \exp^{- \delta / \delta_3}, &  \mbox{ for }  \delta_3< \delta 
 \end{array}
 \right. 
 \nn &&
 \delta_0\approx 10, \, \delta_1 \approx 10^4,   \, \delta_2 \approx 10^6,     \, \delta_3 \geq  10^8
 \label{ppp}
 \ea
(The value of the exponential cut-off is not well determined,  $  10^8 \leq \delta_3\leq 10^{12}$).

As the estimate of the highest energy that a particle can reach, we can balance curvature losses in the \Bf\ with the curvature radius of the order of the \LC, $R_c = \zeta R_{LC}$,  with acceleration by the \Ef\ of the order of the \Bf, \cite{2012ApJ...754...33L},
\be
\gamma_3 \sim  \left( {B_{NS} R_{NS}^3 \over c e P} \right)^{1/4} \,  \left( { E\over B} \right)^{1/4}\,   \left( { R_c \over R_{LC}} \right)^{-1/4} = 10^ 8 \,  \left( { E\over B} \right)^{1/4}\,   \left( { R_c \over R_{LC}} \right)^{-1/4} 
\ee
Note, that the derived $\gamma_3 \approx 2.5 \times 10^8$ (see below) is close to this theoretical limit (for smaller $\delta_3$ the cyclotron spectrum has an upper cut-off below $\sim $ MeV energies.)

Most importantly, we assume that the parallel distribution function 
is constant throughout the \ms\ - an obvious simplifications. This also implies that the motion of particles in the \ms\ is non-adiabatic. For particles gyrating in the dipolar \ms, and in the absence of any other interaction,  there an effective parallel and perpendicular force due to the conservation of the first adiabatic invariant.  The assumption of non-adiabaticity (besides being a highly simplifying in the model) implies (and is justified by the  assumption) that other non-adiabatic forces are at play: particles are accelerated by the parallel \Efs, they produce secondary pairs via various radiative processes, are excited to higher Landau level by absorbing radiation, simultaneously lose their transverse energy by emitting cyclotron photons and, possibly gain transverse energy by emitting photons at the anomalous Doppler resonance, see  \cite{lbm99} and \S \ref{anomalos}.

In addition, we neglected the influence of the curvature and IC radiation reaction on the particle motion. To verify the validity of this assumption let us compare a total electron flux $\dot{N}_e$, normalized to the GJ flux through open field lines, to the total photon flux at GeV energies $\dot{N}_{ph}$:
\ba &&
\dot{N}_e = \lambda n_{GJ} c \pi R_{PC}^2 = \lambda B_{NS} {\Omega^2 R_{NS}^3 \over 2 e c} 
\nn &&
\dot{N}_{ph} = L_\gamma /(\epsilon_\gamma \Delta \Omega)
\nn &&
{\dot{N}_{ph}  \over \dot{N}_e}  \approx {10^6 \over \lambda}
\label{FFF}
\ea
Thus, for $\lambda > \geq 10^6$ each secondary particle emits more than one photon -  radiative drag is then important in changing the parallel momentum (since in the KN regime each scatter results in a particle's momentum change of the order of unity). Since our model predicts $\lambda \sim 10^6$  (see below),  as a simplifying assumption we can neglect radiation reaction on the parallel particle motion.

\subsubsection{Transverse distribution}
Within the CSC model there are two unknown functions: distributions of  parallel and perpendicular momenta. For the perpendicular distribution  we assume that in the center of gyration of the emitting leptons the transverse velocity is non-relativistic, so  in that
frame the photons are emitted at the  local cyclotron frequency.
Then,  the transverse distribution function controls mostly the intensity of the cyclotron emission. In principle the value of the  transverse velocity should be calculated self-consistently using some excitation mechanism, balanced by radiative losses. A possible excitation mechanism is outlined in \S \ref{anomalos}. For now, we just use a parameterization $\beta_0(r)$. We note that the transverse velocity should be negligible below some radius $r_{min}$ (otherwise the cyclotron photon peak frequency will be at too high frequencies). Somewhat arbitrarily we chose  $\beta_0(r)= \beta_0 ( \eta_r-   \eta_{r,min})^3$ (see, though, \S \ref{anomalos} for a possible justification of this scaling). Partly this choice is motivated by simplicity and the post-factum good resulting fit. Also, this choice of the perpendicular velocity is not independent, it is a function of the parallel velocity distribution. 

\subsection{Cyclotron emission}

In \S \ref{outline} we presented simple order-of-magnitude estimates that demonstrate the validity if the CSC model. 
Let us next add  a spread in  parallel momenta and reproduce some of the details of the observed spectrum.  The above relations for  cyclotron emission produced by the mono-energetic beam can be easily generalized to a broad distribution by multiplying the practice density (Eq. (\ref{etatot}) by $f(\delta) d  \delta$ , $ \dot{N} \rightarrow \dot{N}  f(\delta) d  \delta$ and integrating over the Doppler factor $\delta$. (For convenience we express the distribution function in terms of the Doppler factor $\delta$ and not the particle momentum.) 

The cyclotron emissivity for a broad distribution is then
\be
\eta_s = {1 \over 3 \pi} b_{NS}^2 \beta_0^2 \delta \Delta \Omega \eta_\Omega \lambda {c R_{NS}^2 \over \lambda_C^3}
\delta_D\left( \tilde{\epsilon}- b_{NS} \delta/\eta_R^3\right)\, 
 f(\delta) d \delta \,  {d \eta_R \over \eta_R^4} \, d \tilde{\epsilon} 
\ee
Integrating over $\delta$ we find
\be
\eta_s = {1 \over 3 \pi}  \beta_0^2 \Delta \Omega \eta_\Omega    \lambda  {c R_{NS} \over \lambda_C^3}  f\left( { \tilde{\epsilon}  \eta_R^3 \over b_{NS}}\right)\,  \tilde{\epsilon}     d \tilde{\epsilon} \, \eta_R^2   d \eta_R 
\label{sds}
\ee
To proceed further (the integration over radius) we assume a particular distribution function given by Eq. (\ref{ppp})
 The photon spectral density  $n_s(\tilde{\epsilon}, \eta_R)$ at each radius $\eta_R$ is then found by integrating the cyclotron emissivity (\ref{sds}) from this radius to the maximal (\LC) radius.  The total cyclotron flux is then found  by integrating the emissivity (\ref{sds}) from some minimal radius (parameter $\eta_{r, min}$ to be fitted) up to the light cylinder  (parameter $\eta_{r, max} =160$).

\subsection{IC emission} 
After the soft  photon spectral density is found at each radius we calculate the IC emission.
Calculations of the IC emission can be done in a general case, but since the scattering is generally in the deep KN limit, the corresponding relations simply considerably. In this case  the IC photons come out with distribution $ \propto \delta_D( \tilde{\epsilon}_{IC} - \delta /2) f(\delta) d \delta$, where $f(\delta) $ is the distribution function of parallel momenta.  If the density of the soft photons at radius $\eta_R $ is $ n_s (\tilde{\epsilon}_s, \eta_R)  d \tilde{\epsilon}_s$, the IC luminosity is
\be
\eta_{IC} = {b_{NS} \over 4\pi} \Delta\Omega \eta_\Omega \lambda n_s (\tilde{\epsilon}_s, \eta_R)  { d \tilde{\epsilon}_s  \over \tilde{\epsilon}_s }\delta_D ( \delta - 2 \tilde{\epsilon}_{IC}) 
 { c r_E R_{NS}^2 \over \lambda_C}  f(\delta) d\delta \, d \tilde{\epsilon}_{IC} \, {d \eta_R \over \eta_R} 
\ee
Integration over $\delta $ gives
\ba &&
\eta_{IC} = {\cal A}\, n_s (\tilde{\epsilon}_s, \eta_R) \, f(2 \tilde{\epsilon}_{IC}){ d \tilde{\epsilon}_s  \over \tilde{\epsilon}_s } \,  {d \eta_R \over \eta_R}   d  \tilde{\epsilon}_{IC}
\nn &&
 {\cal A}= {b_{NS} \over 8 \pi \tilde{\epsilon}_{IC}  }  \Delta\Omega \eta_\Omega \lambda { c r_E R_{NS}^2 \over \lambda_C}
 \ea
 Coefficient $ {\cal A}$ is the overall normalization, while the shape of the IC spectrum is determined by the particle  distribution function $f$, convolved with the photon density $n_s$,  dependence of the KN cross-section, $\propto 1/\tilde{\epsilon}_s$ and the dependence of the soft photon density on radius. 

\subsection{Fit to data}
\label{fit}
\begin{figure}[h!]
\includegraphics[width=.99\linewidth]{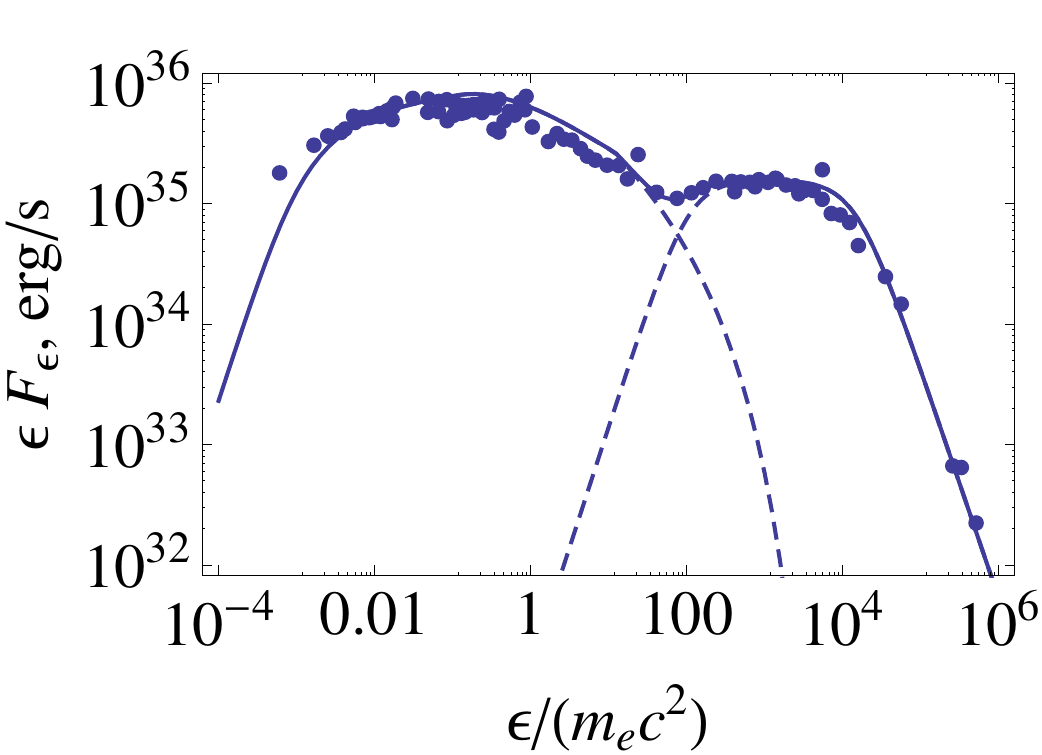}
\caption{The broadband spectrum of the Crab approximated with the CSC model. The data are from \cite{2001A&A...378..918K,2010ApJ...708.1254A,VERITASPSRDetection}.  The IC bump in the KN regime provides a direct measurement of the bulk  particle distribution, while the high energy part of  cyclotron bump constrains the very high energy tail of the particle distribution.
 This is a fit over nearly ten decades in energy, using only a handful of parameters. } 
\label{CrabPulsarFit}
\end{figure}

The model is compared with the data in Fig. (\ref{CrabPulsarFit}). We stress that this is only a comparison and not a multi-parameter fit. Thus, we do not plot the error bars, nor calculate various goodness-of-fit quantities.  The purpose of this figure is not  to make a detailed fit of the spectrum (this would require a more detailed modeling), but to demonstrate that under very general assumption the CSC model reproduces main observational features. For this particular figure the parameters are 
$\delta_0 =1.5$ (the minimal energy cut-off; the value  of $\delta_0 $ determines the low rise energy of the IC component), $m=1$ (particle energy spectrum below the first break; $m$ is determined by the IC spectrum below the break, in the \Fermi\ band, \citep{2010ApJ...708.1254A}),   $\delta_1= 3 \times 10^4$ (the first particle break energy), 
$n=-1$  (particle energy spectrum above the first break),   $\delta_2= 5 \times 10^6$ (the second particle break energy; both $n$ and  $\delta_2$ determine the higher energy part  of the cyclotron SED), 
$l=-2.9$  (particle energy spectrum above the second break; $l$ is most sensitive to the IC spectrum above the break determined by  \cite{VERITASPSRDetection}),  $\delta_3= 5 \times 10^8$  (the exponential cut-off), $\eta_{r,min}= 20$ (the minimal distance from the star where cyclotron emission is produced, $\lambda = 5.5 \times  10^6$ (average particle multiplicity; $\lambda$ determines the relative scaling of the cyclotron and IC component), $\beta_0= 9 \times 10^{-10} ( \eta_{r}-\eta_{r,min})^3$ 
(the transverse velocity in the center of momentum frame; $\beta_0$ determines the  overall normalization, scaling with radius,  chosen by the prediction of the balance between quasilinear diffusion at the anomalous resonance and spontaneous emission at the normal cyclotron resonance, see \S \ref{anomalos},   mostly determines the shape and the  relative intensities of the low and high energy tails of the cyclotron component). 
(The small value of $\delta_0 $ formally violate our assumption of the IC scattering in the deep KN regime, but we deem it acceptable giving the simplicity  of the model and the uncertainties of the multi-parameter fit.)

The required plasma multiplicity, $\lambda \sim 10^6-10^7$, is higher than is typically achieved in pair production models \citep{AronsScharlemann,2001ApJ...554..624H,2011ApJ...743..181H}, but is comparable  to the  {\it average} multiplicity of $\lambda \sim 10^6$ required by the observations of   the Crab nebula's $X$-ray emission  \citep{1970ApJ...159L..77S,1996A&AS..120C..49A}, see also \S \ref{twophot}.

\section{Two photon pair production in  outer gaps}
\label{twophot}

The models of the pulsar \mss\ are mostly based on the  \cite{1975ApJ...196...51R} model of pair production near the polar caps. It was then expected that polar cap regions are intense sources of high energy emission \citep{1982ApJ...252..337D,1996ApJ...458..278D,HardingMuslimov98,2000ApJ...532.1150Z,2004AdSpR..33..552B}. These expectations are not supported by few years of \Fermi\ data,  no gamma-ray emission from pulsar polar caps have been observed. This calls into question the very paradigm of polar cap particle creation. Previously, pair production in the outer gaps between the $\gamma$-ray photons and surface $X$-ray photons were considered by \cite{2000ApJ...537..964C}.

On the other hand,  the Crab nebula $X$-ray emission requires a huge flux of particles, with the {\it average} over the open field lines multiplicity of $<\lambda> \sim 10^6$ \citep{1970ApJ...159L..77S}. Thus, there is a  contradiction between polar cap pair production models and observations. The current model provides an interesting new possibility for pair production. IC scattering in the KN regime is tightly related to the two photon pair production: the corresponding cross-sections are very similar \citep[\eg][]{1970RvMP...42..237B,2004vhec.book.....A,2009herb.book.....D}.

The two photon pair production on photons of energy $\tilde{ \epsilon}_1 $ and $\tilde{ \epsilon}_2 $ is a threshold process, which requires $s_0 =\tilde{ \epsilon}_1\tilde{ \epsilon}_2 >1$. For a soft photon produced at region of \Bf\ $b$ (in terms of quantum \Bf) and having reduced energy in the observer frame $\tilde{ \epsilon}_1= \delta b$,  interacting head-on with the  IC photon scattered in the KN regime, $\tilde{ \epsilon}_2 \approx b/2$, the threshold condition reads
\be
b \delta^2/2 = 1 \rightarrow \eta_{R}= (b_{NS}/2)^{1/3} \delta^{2/3}   \approx 100
\label{ee}
\ee
for $\delta \approx  2 \tilde{\epsilon_{IC}} \approx 4000$.
Thus,  the threshold conditions for two photon pair production are satisfied nearly in the whole \ms\ of Crab pulsar (recall that in Crab the \LC\ is located  at $\eta _R=160$ and both the soft and the VHE photons have a broad spectral distribution).

The pair-production cross-section has a maximum at the level of $\sigma_{\gamma \gamma} \approx  0.2 \sigma_T$  achieved at $s_0 \approx  3.5 - 4$ \citep{2004vhec.book.....A}. For $s_0 \gg 1$,
\be
\sigma_{\gamma \gamma} \approx (2/3)  \sigma_T {\ln s_0 \over s_0}
\ee
Estimating the  photon density as $ n_{\rm ph} \sim {L_X /( \Delta \Omega r^2 \tilde{\epsilon}_{IC} m_e c^2)}$, 
the optical depth for the  two photon pair production  at the location (\ref{ee})is very high:
\be
\tau _{\gamma \gamma}\approx r \sigma_{\gamma \gamma} n_{\rm ph} 
 \approx {3 \over 2^{5/3}} { L_x \delta^{1/3} \sigma_T \over b_{NS}^{1/3} c^3 m_e R_{NS} \Delta \Omega} = 10^{4} \Delta \Omega_2
\ee
Such a very high optical depth is, naturally, an overestimation, since then no $\gamma$-ray emission would then be expected \citep[\cf the model of gap closure 
by the photon-photon pair-creation process between the high-energy $\gamma$-rays emitted in the gap and the $X$-rays coming from
the stellar surface by][]{1997ApJ...487..370Z}. For example, in the current estimate 
we assumed that all the soft and VHE photons interact in head-on collisions at a peak of pair production cross-section - clearly an upper limit. Still,  a very large value of $\tau_{\gamma \gamma}$ implies that a $\gamma-\gamma$ pair production between the soft and VHE photons is an important process in pulsar \mss.

The possibility that outer gaps are the main sources of pair leads to the requirement of the very high particle density. If the average  over the open field lines multiplicity is $<\lambda> \sim 10^6$, since the gaps occupy a small fraction of the open field lines solid angle, the  multiplicity within the gaps should be $
\lambda \sim <\lambda> \pi \eta_\Omega/\Delta \Omega \approx 2  <\lambda> \, \Delta\Omega_{-2} \eta_R$; \ie tens to hundreds times higher. This compares favorably with our independent  estimates, \S  \ref{fit}.

To construct a self-consistent model of pair production in the outer gaps is a formidable non-linear problem:  the plasma density depends on the photon field, which in turn depends on the plasma density;  the accelerating \Ef\ also depends on the plasma density; the excitation of  particle gyration, which controls the cyclotron emission rates,  also depends on the details of the distribution function (both parallel and transverse, via the growth rate of the maser instability), while the evolution of the distribution function  also  depends on the state of particle gyration, see next Section.

\section{Relating radio to high energy  emission}
 \label{anomalos}

The present model of pulsar high energy  emission requires that the emitting particles have a finite pitch angle. This requires a mechanism that would excite particle gyration. In this Section we outline  such a model, to be addressed in more details in a subsequent publication.

A possible excitation mechanism is  related to the generation of the pulsar radio emission at the anomalous cyclotron resonance \citep{1977ApJ...217..832K,mu79,Kaz91,lbm99}. In this model
the pulsar radio emission is  produced directly
 by   maser-type
plasma instabilities operating at the anomalous cyclotron-Cherenkov resonance
$\, \omega-\, k_{\parallel} v_{\parallel} +\, |\omega_B|/\,  \gamma_{res}=0$ (note the sign in front of the cyclotron term).
The
instabilities are due to the interaction of the fast particles from
the primary beam and  the tail of the distribution
 with the  normal modes of a strongly magnetized one-dimensional
electron-positron plasma.
The waves emitted at these resonances are vacuum-like,  electromagnetic
waves that may leave the  magnetosphere directly
\citep{1999ApJ...512..804L}.

At the anomalous resonance a particle emitting a wave undergoes a transition up in Landau levels. Thus,  initially one-dimensional distribution develops a finite pitch angle. As a result, resonant particles start emitting  cyclotron photons at the normal cyclotron resonance. The transverse particle distribution can achieve a balance between the diffusive spread due to the coherent interaction with the waves at the anomalous resonance and spontaneous photon emission at the normal resonance \citep{Kaz91,2011ApJ...730...62C,LyutikovQL,2011ApJ...730...62C}. In our notation, the typical transverse momentum \citep[Eq. (11) of][]{2011ApJ...730...62C} is
\be
\beta_0 \approx {1\over b_{NS}} \, {\lambda_C \over (r_E r_{NS}^3)^{1/4}} \, \eta_{R}^3 \, \eta_\Omega^{3/4}
\label{beta0}
\ee
The scaling with radius is the one chosen for the fit, \S \ref{fit}. Typical values of $\beta_0$ derived using quasilinear diffusion, Eq. (\ref{beta0}), are of the same order of magnitude as required by the spectral fit, \S \ref{fit}.

 In addition, other radiative processes will affect the particle distribution function: (i) excitation of cyclotron motion due to IC scattering and (ii) during two photon pair production; (iii) induced Raman scattering within the \ms, \citep{1998MNRAS.298.1198L}; (iv) single particle cyclotron absorption \citep{2000A&A...355.1168P}. Since these processes can be nonlocal within the magnetosphere (a photon emitted at one location can be absorbed/scattered at  a very different location), this presents a complicated electrodynamic problem. 

In this Section we discussed how excitation of coherent radio  wave at the anomalous cyclotron resonance in the outer gaps can excite the cyclotron motion of the resonant particles. Spontaneous emission of the cyclotron photons then produces the UV-X-ray bump, while IC scattering produces the VHE emission. Due to the highly relativistic motion all the emission components are beamed along the local \Bf: this explains the similar profiles of Crab pulsar from radio to VHE gamma-ray emission. Thus, {\it  the model unites the Crab pulsar non-thermal emission from tens megahertz to hundreds of GeVs, nearly eighteen decades in energy.}
Note that the outlined model, uniting radio and gamma ray emission, is applicable specifically to Crab pulsar, and specifically to  the main pulse and the interpulse, and not the radio precursor. Recall, that Crab is one of the few pulsars where radio and high energy peaks are nearly aligned in phase. In most pulsars, radio peak precedes the gamma rays peaks \citep{2010ApJS..187..460A}. A relatively weak radio precursor in Crab is commonly associated with the "normal" radio emission, while the main pulse and the interpulse are, probably,  "different" emission mechanisms \citep[\eg][]{hankins96}.

\section{Conclusion}

In this paper we demonstrate that the cyclotron-self-Compton model  is a viable model of pulsar high energy emission. Using   observationally-constrained  properties of the distribution function   of particles within the pulsar \ms, as well as  fairly general additional theoretical assumptions, we are able to reproduce the overall SED of the Crab pulsar over nearly ten decades of energy. The key theoretical assumption is the presence of two counter-streaming populations within the outer gaps. The outward moving beam then produces UV-$X$-ray photons via Doppler-booster cyclotron emission, and the GeV emission by Compton scattering the cyclotron photons produced by the inward going beam. As a simplifying assumption, the parameters of the inward and outward going beams were chosen to be the same: if the distribution is skewed in favor of the outgoing beam, the intensity of the ingoing radiation will be reduced, producing a nearly one-sided emission pattern (in addition, cyclotron absorption may be important for the inward going radiation \citep{1995JApA...16..173R,2000ApJ...537..964C}).

Naturally, the current one-dimensional model is a simplification; it is expected that more advanced models, that incorporate self-consistently the geometry, structure of accelerating electric fields and various radiative processes,  will result in the modification of the inferred parameters. The most important simplifications include,  first, the assumption of a IC scattering in a deep KN regime is not valid for scattering of the cyclotron photon emitted close to the \LC\ by the slowly moving particles. Second, the separation of the distribution function into parallel and perpendicular parts is likely not to be a unique from the observation point of view. For example, relativistic transverse motion vill increase the emitted frequency in the center-of-gyration frame, thus reducing the required parallel boost.

I would  like to thank Jonathan Arons, Charles Dermer, Kouichi Hirotani and George Machabeli for encouraging discussions.

\bibliographystyle{apj}
  \bibliography{/Users/maxim/Home/Research/BibTex} 

\end{document}